\documentclass[doublecol,linenumbers]{epl2}

\usepackage{bm}
\usepackage{graphicx}
\usepackage{multirow}

\usepackage{float}

\title{Network reconstruction by the stationary distribution of random walk process}
\author{Zhe He\inst{1} \and Ming Li\inst{1,2} \thanks{E-mail: \email{minglichn@ustc.edu.cn}} \and Rui-Jie Xu\inst{1} \and Bing-Hong Wang\inst{1}\thanks{E-mail: \email{bhwang@ustc.edu.cn}}}
\shortauthor{Zhe He \etal}
\institute{\inst{1} Department of Modern Physics, University of Science and Technology of China - Hefei, 230026, P. R. China \\
\inst{2} Department of Applied Physics, Hong Kong Polytechnic University - Hung Hom, Hong Kong
}


\abstract{It is known that the stationary distribution of the random walk process is dependent on the structure of the network. This could provide us a solution of the network reconstruction. However, the stationary distribution of the random walk process can only reflect the relative size of node degrees directly, how to infer the real connection is still a problem. In this paper, we will propose a method to reconstruct network by the random walk process, which can reconstruct the total number of links, degree sequence and links sequentially. In our method, only the stationary distribution is used, and no data of the evolution process is needed, such as the first passage time. We perform our method on some network models and real-world network, the results indicate our method can reconstruct networks accurately, even when we can not get the exact stationary distribution.}

\begin{document}
\maketitle

\section{Introduction}

Many real-world complex systems can be considered as complex networks, such as biology\cite{x1,x3}, psychology\cite{x4}, social and economic systems\cite{x2}. The dynamics in these networks are generally determined by their structures, so revealing the structure of network systems is one of the important ways to study the dynamics of complex systems\cite{Boccaletti2006175}. However, not all the structures can be detected directly in practice, such as some biology network systems. To obtain the structure of such systems, the data-driven reconstruction is usually used\cite{x5,guimera2009missing}. Generally speaking, that is inferring the connection of an unknown network by analyzing the feedback information of some dynamics in that system.

Previous studies have already obtain many significant results with various reconstruction techniques. These reconstruction techniques are often based on data analysis of the time series of network dynamics, and the widely used dynamics are the ones that can be expressed as some first-order non-linear differential equations. For example, Timme \emph{et al} reveal network connectivity by response dynamics\cite{x8}, Wang \emph{et al} infer the network structure with noise-bridge dynamics\cite{x9}, Levnajic \emph{et al} reconstruct network from random phase-resetting\cite{x10}, and Yu \emph{et al} estimate topology of networks with phase oscillators\cite{x11}. Besides, some special dynamics are also used in biology studies\cite{x12,x13,x6,x7,x18}, such as DBNs\cite{x14} and gene regulatory\cite{x19}. Recent works extend reconstruction techniques to game theory\cite{x16}, mean field theory\cite{x15}, compressed sensing\cite{x17}, epidemic spreading\cite{altarelli2014bayesian,liao2015reconstructing} and
many other techniques\cite{x20,x21,x22}. Even in noisy network observations, some method can also be used to estimate the true network properties\cite{Guimera29122009}. In addition, the link prediction methods are also one kind of the network reconstruction, which are usually used in recommender systems\cite{Lu20111150,Lv20121}. A difference is that the link prediction methods often use the information of the known links to infer the missing ones, however, the dynamic data is usually used for the traditional network reconstruction

As a basic conclusion of the Markov process on networks, the stationary distribution of the random walk process depends on the network structure\cite{PhysRevLett.92.118701}. This correlation of the network structure and dynamic could also provide us a method of the network reconstruction. However, this correlation can only reflect the relative size of node degrees, how to infer the real connection is still a problem. In this paper, we will propose a method based on random walk process to infer the network connection. Only the stationary distribution is used in our method, the information of the evolution of random walk process is not required, such as the first passage time. So our method is more simple and fundamentally different with the one used in ref.\cite{Wittmann20093826}.

This paper is organized as follows. In the next section, we review some basic conclusion of random walk on network first, and then give the details of our network reconstruction method. In Sec. III we will perform our method on some network models and real-world network. In the last section we will give a simple discussion.

\section{Method}

\emph{random walk}- The random walk process begins with a walker on a node, labeled as $i$. Then, it walks to one of its neighbor $j$ with probability $p_{ij}$, the so-called transition probability. If $p_{ij}=1/k_i$, where $k_i$ is the degree of node $i$, this process is a totally random walk process, i.e., the walker has the same probability to go to any of its neighbors. For the case $p_{ij}\neq 1/k_i$, it is a biased random walk. However, the following conditions must be always met, $1)$ $p_{ij}=0$, if nodes $i$ and $j$ are disconnected. $2)$ $\sum_jp_{ij}=1$, where the sum over all the neighbors of node $i$.

In a connected network with size $N$, assuming there is only one walker, our focus is the probability $\pi_i(t)$ that the walker is on node $i$ at time $t$. Then, the master equation for the probability $\pi_i(t)$ can be written as
\begin{equation}
\pi_i(t+1)=\sum_jp_{ji}\pi_j(t). \label{pit}
\end{equation}
If $p_{ij}=1/k_i$, when $t\rightarrow\infty$, i.e., in the steady state, it is well known that\cite{PhysRevLett.92.118701}
\begin{equation}
\pi_i=\frac{k_i}{\sum_ik_i}=\frac{k_i}{2L}. \label{pi}
\end{equation}
Here, $L$ is the total number of links in the network. This is the stationary distribution of the random walk process.

\emph{reconstruct degree}- Obviously, eq.(\ref{pi}) can be used to infer the degree of each node of a network, when we obtain the stationary distribution $\pi_i$ of the random walk process in the network. However, the total number of links $L$ is also an unknown parameter, so we need to infer $L$ firstly. Once the total link number is obtained, we can infer the degree sequence by eq.(\ref{pi}) directly. Note that the number of nodes $N$ is a known parameter.

Next, we will show how to infer the total number of links. For the real total number of links $L_0$, all the degrees obtained by eq.(\ref{pi}) will be integers. However, a bad choice of $L$ in eq.(\ref{pi}) could make almost all the degrees be non-integers. Therefore, we can use the following parameter to quantitative analysis,
\begin{equation}
\Delta_l=\frac{\sum_i|k_i^l-\lfloor k_i^l+0.5\rfloor|}{N}, \label{delta}
\end{equation}
where $k_i^l$ is the degree of node $i$ obtained by letting $L=l$ in eq.(\ref{pi}). Note that $k_i^l$ may not be an integer. So, an integer degree we get from eq.(\ref{pi}) can be written as $\lfloor k_i^l+0.5\rfloor$, and $\lfloor x\rfloor$ gives the largest integer that smaller than $x$. Considering the statistics fluctuation of $\pi_i$, it is clear that the $l$ corresponding to the minimum of $\Delta_l$ is the real total number of links $L_0$.

In practice, a rigorous way to find $L_0$ is checking $\Delta_l$ for every $l$ from $l=N-1$\cite{note}. In fact, however, we need not to check so many $\Delta_l$ to find $L_0$. The reason is as following. As indicated by eq.(\ref{delta}), the minimum of $\Delta_l$ can also be found at $l=nL_0 (n=2,3,4,...)$. Thus, if we find two nearby minimums at $l_1$ and $l_2$ ($l_2>l_1$) that satisfy $l_1=l_2-l_1$, we can determine $l_1$ is the total number of links $L_0$, immediately. In addition, due to the statistics fluctuation of $\pi_i$ in practice, the minimums at $l=nL_0$ for large $n$ could be larger than that of $l=L_0$. Our simulation results in the next section will confirm these.

More important, this method of inferring the total number of links depends on that all the real degrees have no common factors. In a large network, it normally can find more than one node with prime-number degrees, which will make all the degrees have no common factors. Therefore, our method can be used in most of networks to infer the total number of links.

\emph{reconstruct connection}- As we know, the networks with the same degree distribution but different correlation could demonstrate entirely different phenomenon. So reconstructing the connection is also important after the degree sequence is obtained.

Since the random walk process is a Markov process, the stationary distributions $\pi_i$ and $\pi_j$ of two connected nodes $i$ and $j$ is positive correlation. That is when a node is visited with a high frequency by the walker, its neighbors will also be visited frequently. Correspondingly, if nodes $i$ and $j$ are non-adjacent, $\pi_i$ and $\pi_j$ will be independent of each other. This provides us a method to infer the connection of the network. 

First, we perform the random walk process $M$ times on the network, thus for each $\pi_i$, we will have $M$ samples. Second, calculating the covariance $cov(\pi_i,\pi_j)$ of the $M$ samples for each pair $\pi_i$ and $\pi_j$. At last, connecting node pairs with the largest covariance one by one, until the degree sequence has been satisfied. In this step, when the degrees of two nodes are satisfied, they no longer need to be considered, even if the corresponding covariance is larger.

It may be added, in this step, there is no need to fix the transition probability $p_{ij}=1/k_i$ for each random walk process, since this correlation of the stationary distributions $\pi_i$ and $\pi_j$ is also satisfied for biased random walk. This means that we need not to know the details of the transition probability for each realization. However, in order to reconstruct degrees by eq.(\ref{pi}) as shown in the last step, the transition probability $p_{ij}$ for each realization must be chosen from a distribution with expectation $1/k_i$. In this way, we can use the average $\overline{\pi_i}$ of these $M$ random walk processes to obtain the degrees, and then use the covariance $cov(\pi_i,\pi_j)$ of the $M$ samples to infer the connection.

To summarize, we can reconstruct a network as followings.
\begin{enumerate}
  \item Performing the random walk process $M$ times, for which the transition probability $p_{ij}$ is chosen from a distribution with expectation $1/k_i$, randomly.
  \item Using eqs.(\ref{pi}), (\ref{delta}) and the average stationary distribution $\overline{\pi_i}$ of the $M$ samples to obtain the total number of links $L$ and the degree sequence $k_i$.
  \item Calculating the covariance $cov(\pi_i,\pi_j)$ of the $M$ samples for each pair of nodes $i$ and $j$, then connecting node pairs with the largest covariance one by one until all the degrees have been satisfied.
\end{enumerate}

\section{Simulation results}

\begin{figure}\centering
\includegraphics[width=0.48\textwidth]{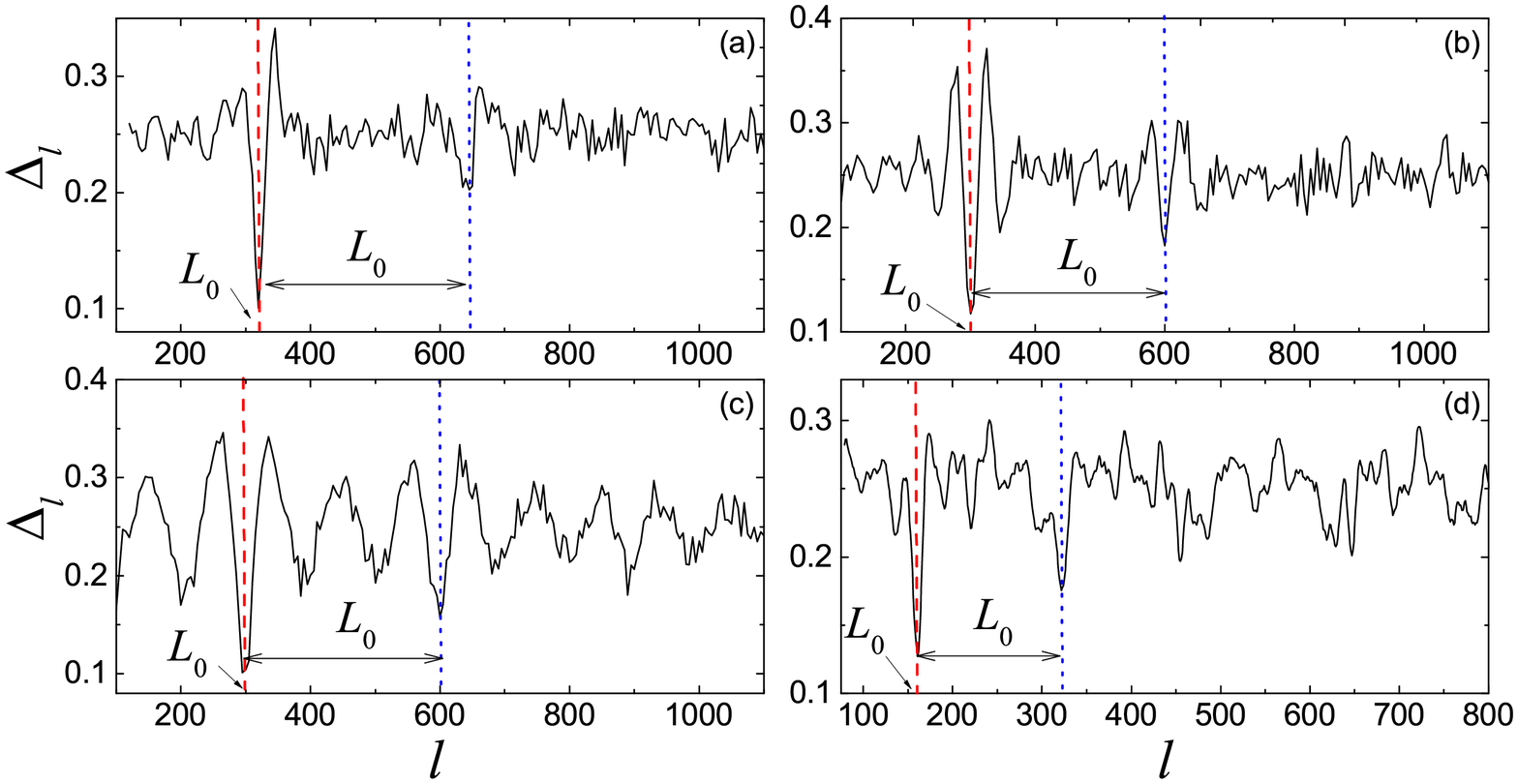}
\caption{(color online) Inferring the total number of links. (a) ER network with connected probability $0.06$. The total number of links of the network shown in this figure is $L_0=320$. (b) WS network with rewiring probability $0.5$. The total number of links is $L_0=300$. (c) BA network with average degree $\langle k\rangle=6$, i.e., $L_0=300$. (d) The dolphin network with $N=62$ and $L_0=159$\cite{LusseauS477}. The sizes of the networks in (a), (b) and (c) are $N=100$. The red dash lines and blue dot lines are labeled the positions of the minimum of $\Delta_l$ at $l=L_0$ and $l=2L_0$, respectively.}
\label{f1}
\end{figure}

\begin{figure}\centering
\includegraphics[width=0.48\textwidth]{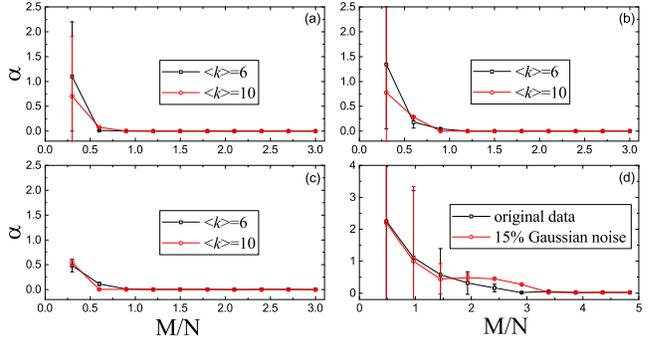}
\caption{(color online) Effect of the sample number $M$ on the accuracy of the link number prediction. (a) ER networks with connected probabilities $p=0.06$ and $0.1$. (b) WS networks with rewiring probability $0.5$ for average degrees $\langle k\rangle=6$ and $10$. (c) BA networks with average degrees $\langle k\rangle=6$ and $10$. (d) The dolphin network. The original date is the average stationary distribution $\overline{\pi_i}$, and the $15\%$ Gaussian noise means that a noise is added in the average stationary distribution $\overline{\pi_i}$, which obeys Gaussian distribution with the expectation $0$ and standard deviation $15\%$. The sizes of the networks in (a), (b) and (c) are $N=100$. All the plots are averaged over $30$ realizations. }
\label{f2}
\end{figure}

First of all, it is needed to point out that in our simulation, the transition probability $p_{ij}$ used in each realization is chosen randomly from a uniform distribution with expectation $1/k_i$. This means that we do not know the detail process of each realization.

To evaluate the accuracy of our method, we first test our method of inferring the total number of links. In Fig.\ref{f1}, we show $\Delta_l$ obtained by our method for Erd\H{o}s-R\'{e}nyi (ER), Watts-Strogatz (WS), Barab\'{a}si-Albert (BA) and dolphin networks\cite{LusseauS477}. One can find that our method gives a good reconstruction of the link number. The stationary distribution $\overline{\pi_i}$ used in Fig.\ref{f1} is averaged from a larger number of samples. If we fix the transition probability $p_{ij}=1/k_i$, only one realization can give a precise link number. In addition, we can find that for $l=nL_0 (n=2,3,4,...)$, $\Delta_l$ also gives a minimum. This is consistent with the conclusions of analysis in the last section. In Fig.\ref{f1} (c), one can find that there are some other periodic minimums besides the ones at $l=nL_0$. These minimums are easy to be ruled out when we want to find the real $L_0$, since the period of the minimums can not be less than $N-1$. In short, the easy way to find $L_0$ is to get the position of the smallest $\Delta_l$, but sometimes it needs more computation.

In Fig.\ref{f2}, we show the effect of the sample number $M$ on the accuracy of the link number prediction, in which $\alpha$ is defined as
\begin{equation}
\alpha=\frac{|L-L_0|}{L_0}. \label{alpha}
\end{equation}
Here, $L$ is the number of links obtained by our method. It is easy to know that the average stationary distribution $\overline{\pi_i}$ could not obey eq.(\ref{pi}) exactly for small $M$. So we can obtain an accuracy link number only for a larger $M$. Furthermore, to check the dependency of our method on the accuracy of the feedback information, we added a Gaussian noise to the average stationary distribution $\overline{\pi_i}$, before we use it to infer the link number. Here, the Gaussian noise means that the noise obeys the Gaussian distribution. We find that this noise has little effect on the reconstruction accuracy.

\begin{figure}\centering
\includegraphics[width=0.48\textwidth]{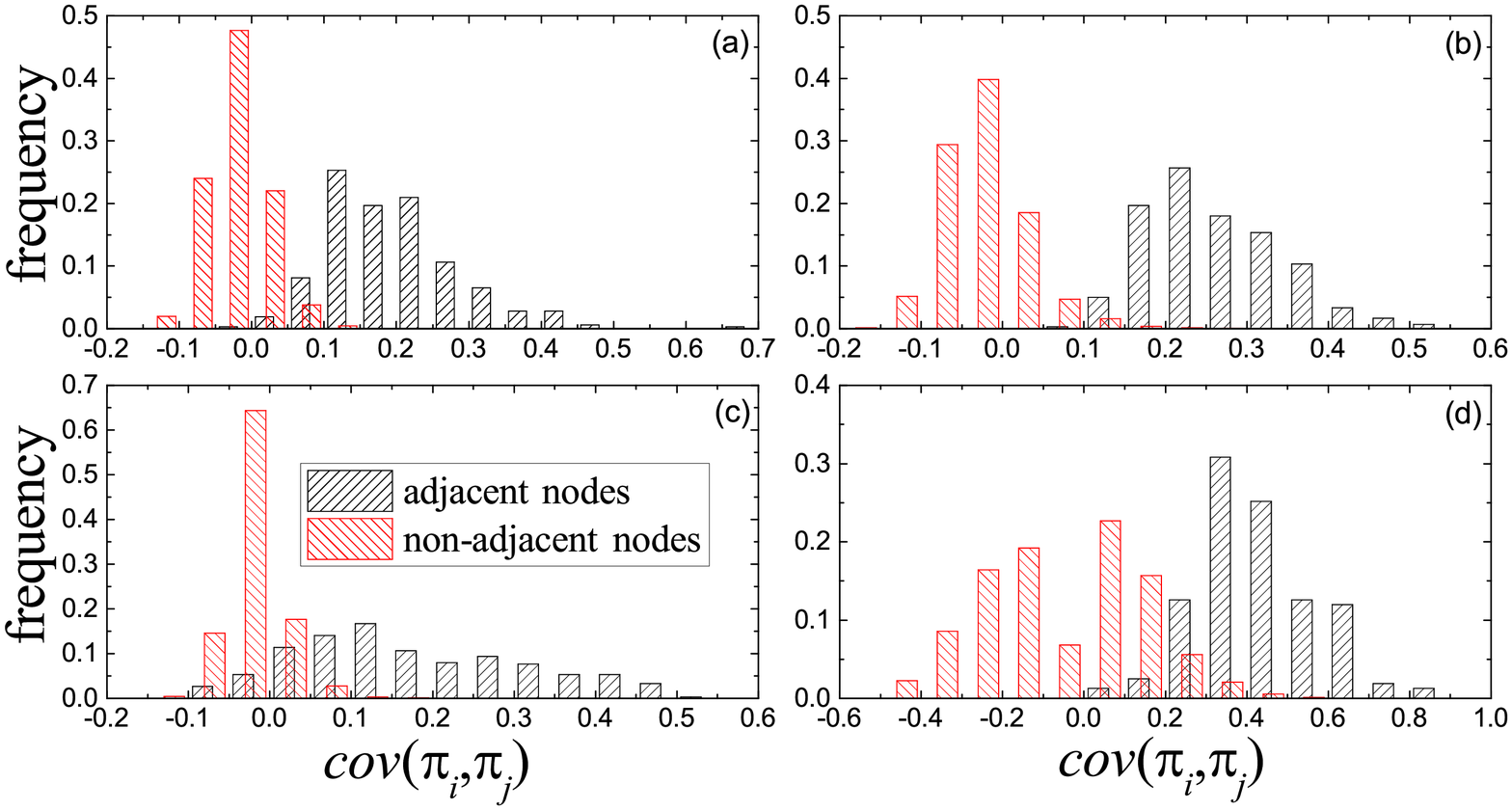}
\caption{(color online) The frequency distribution of the covariance $cov(\pi_i,\pi_j)$ for adjacent and non-adjacent nodes. The four sub-figures are correspond to the ones shown in Fig.\ref{f1}, respectively. The bin size is $0.05$ for (a), (b) and (c), and $0.1$ for (d).}
\label{f3}
\end{figure}

\begin{figure}\centering
\includegraphics[width=0.48\textwidth]{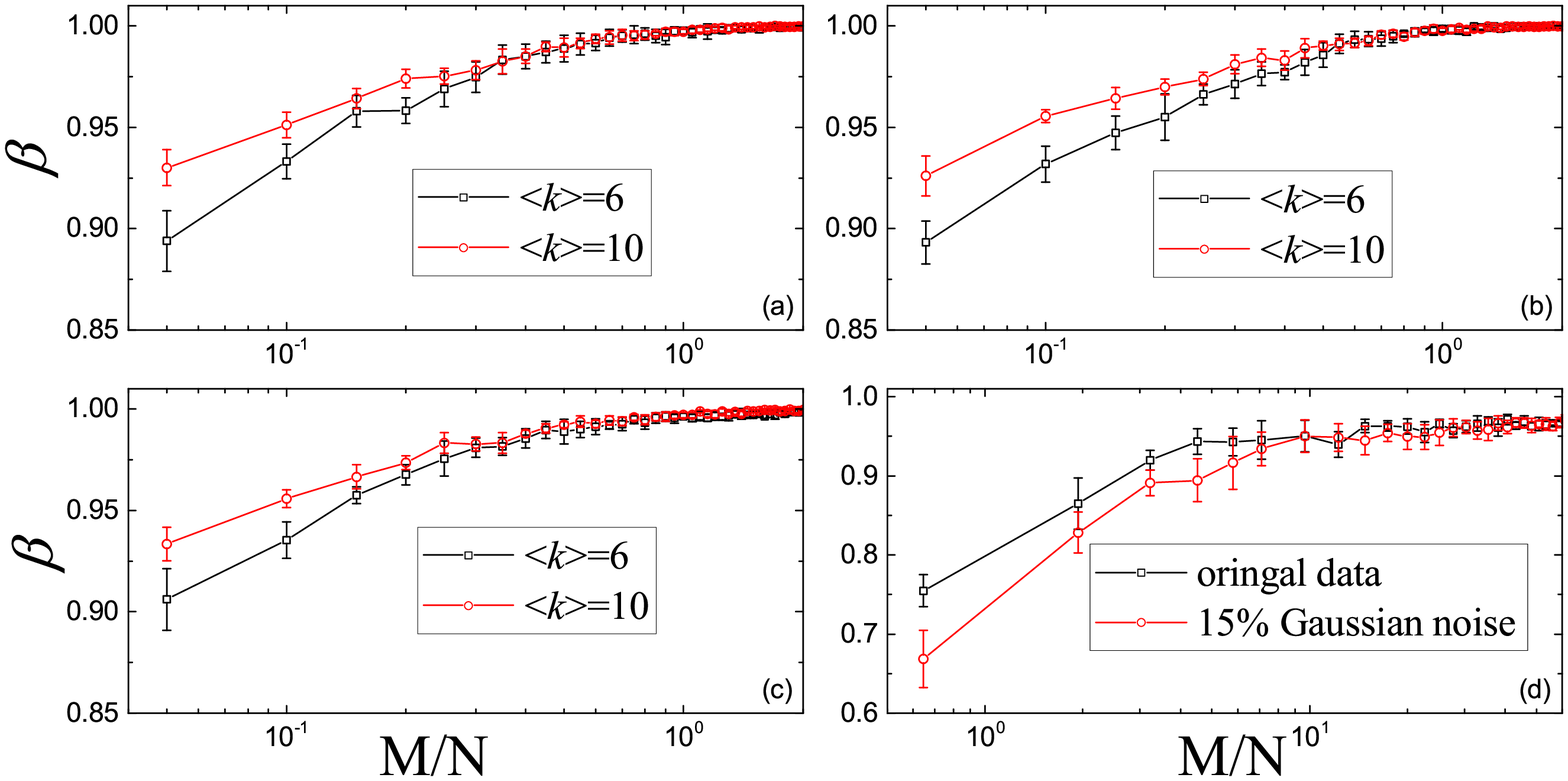}
\caption{(color online) The precision of the degree reconstruction. Each plot is averaged by $30$ realizations. The four sub-figures are correspond to the ones shown in Fig.\ref{f2}, respectively.}
\label{f4}
\end{figure}

\begin{figure}\centering
\includegraphics[width=0.48\textwidth]{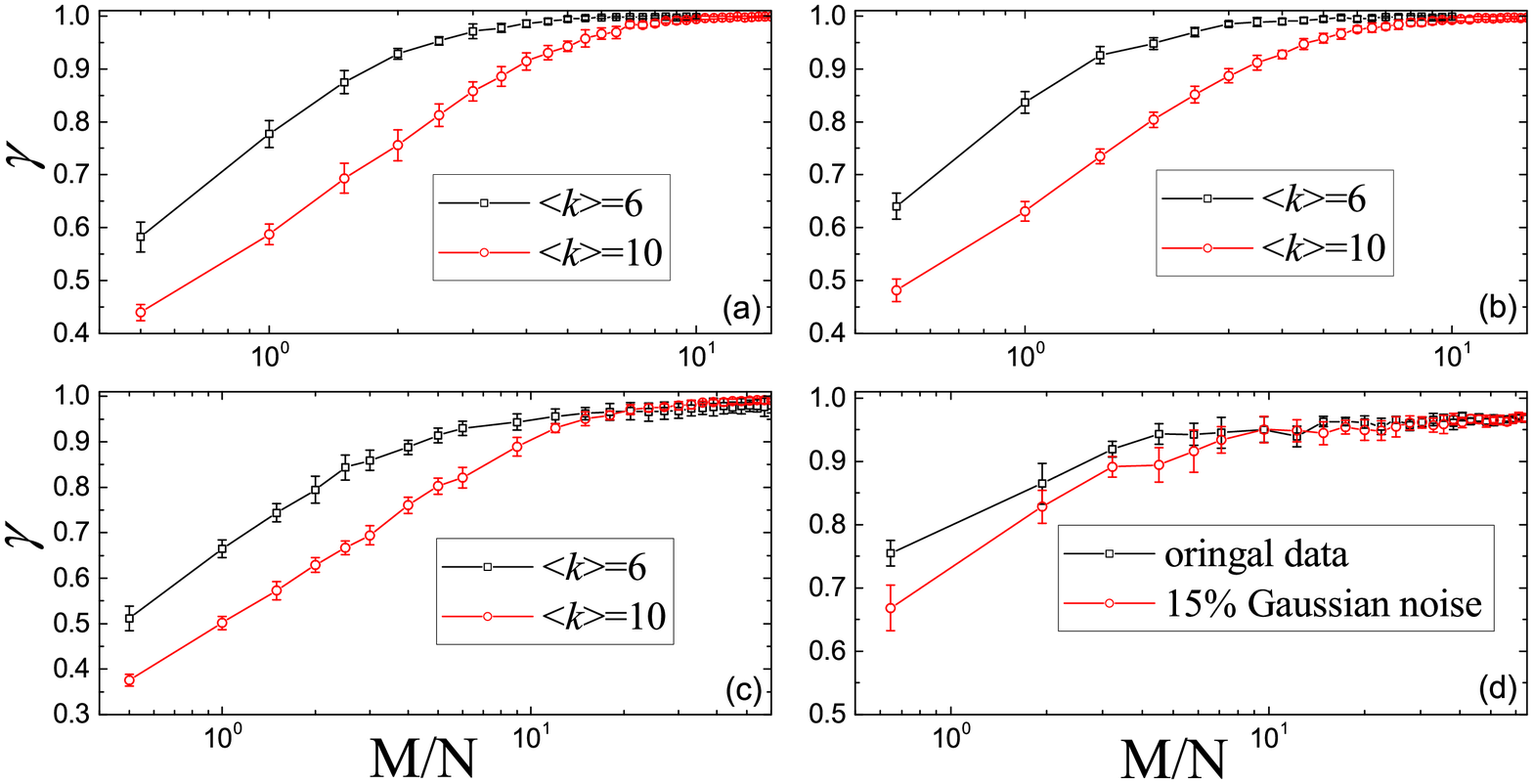}
\caption{(color online) The precision of the link reconstruction. Each plot is averaged by $30$ realizations. The four sub-figures are correspond to the ones shown in Fig.\ref{f2}, respectively.}
\label{f5}
\end{figure}

Before we verify the connection reconstruction, let us check the distribution of the covariance $cov(\pi_i,\pi_j)$ for adjacent and non-adjacent nodes firstly. From Fig.\ref{f3}, we can find that the covariances for adjacent nodes are distributed around a positive value, which means they are correlated. Conversely, those of the non-adjacent nodes are distributed around zero. These results indicate that our method for reconstructing the connection is feasible.

In Figs.\ref{f4} and \ref{f5}, the precision of the links reconstruction is shown. The two parameters use in Figs.\ref{f4} and \ref{f5} are defined as
\begin{eqnarray}
\beta &=& 1-\frac{\sum_i|k_i-k_{i0}|}{2L_0}, \\
\gamma  &=& \frac{L^*}{L_0}.
\end{eqnarray}
Here, $k_{i0}$ is the real degree of node $i$, and $L^*$ is the number of links recovered correctly. We can see that as the sample number $M$ increasing, the precision of the degrees and links reconstruction are both close to $1$. In addition, the reconstruction of dolphin network is not so good as the other three. This is because this network is a high clustered network, and some nodes are hard to distinguish by the stationary distribution of random walk, which results in some erroneous predictions.

\section{Discussion}

Generally speaking, network reconstruction is that through analyzing the relation between the inputs and the corresponding outputs of a black box, and then inferring what is in the black box. For our method, it is not necessary to know the exact inputs (transition probability) for reach realization, we can infer the network structure just by the statistical analysis of the outputs. At the same time, the information of the dynamic evolution process is also not required. These could be helpful for the network system that the inputs and the evolution can not be controlled exactly for each realization.

Actually, this method does not depend on random walk process. A network dynamic with the similar evolution form $X(t+1)=\mathbf{A}X(t)$ could be used to network reconstruction as the way demonstrated in this paper, where $X(t)$ is the state distribution of each node at time step $t$. This is because the matrix $\mathbf{A}$ contains the information of the degree sequence, such as eq.(\ref{pi}). Although for some dynamics, $X$ may not have the simple relation with the degree sequence as eq.(\ref{pi}), with some approximation or mathematical manipulation, this method is still valid. For example, the SIS epidemic model can be solved from an eigenvalue viewpoint\cite{wang2003epidemic}.

Another potential application of reconstructing the degree sequence in our method could be the detection of the hidden nodes. If one gets a network but can not be sure whether it contains all the nodes, we may detect the hidden nodes by comparing the reconstructed degree sequence and the known nodes. One may further infer which nodes connect the hidden nodes.

\acknowledgments
This work is funded by the National Natural Science Foundation of China (Grant Nos.: 11275186 and 61503355) and the Open Funding Programme of Joint Laboratory of Flight Vehicle Ocean-based Measurement and Control under Grant No. FOM2014OF001. ML is also supported by the Fundamental Research Fund for the Central Universities.

\bibliography{ref}

\begin{thebibliography}{10}
\expandafter\ifx\csname url\endcsname\relax\def\url#1{\texttt{#1}}\fi

\bibitem{x1}
\Name{Barabasi A.-L. \and Oltvai Z.~N.} \REVIEW{Nature Reviews
  Genetics}{5}{2004}{101}.

\bibitem{x3}
\Name{De~Jong H.} \REVIEW{Journal of Computational Biology}{9}{2002}{67}.

\bibitem{x4}
\Name{van Geert P. L.~C. \and Steenbeek H.~W.} \REVIEW{Behavioral and Brain
  Sciences}{33}{2010}{174}.

\bibitem{x2}
\Name{Jackson M.~O. \and Watts A.} \REVIEW{Journal of Economic
  Theory}{106}{2002}{265}.

\bibitem{Boccaletti2006175}
\Name{Boccaletti S., Latora V., Moreno Y., Chavez M. \and Hwang D.-U.}
  \REVIEW{Phys. Rep.}{424}{2006}{175}.

\bibitem{x5}
\Name{Hecker M., Lambeck S., Toepfer S., Van~Someren E. \and Guthke R.}
  \REVIEW{Biosystems}{96}{2009}{86}.

\bibitem{guimera2009missing}
\Name{Guimer{\`a} R. \and Sales-Pardo M.} \REVIEW{Proc. Natl. Acad. Sci.
  USA}{106}{2009}{22073}.

\bibitem{x8}
\Name{Timme M.} \REVIEW{Phys. Rev. Lett.}{98}{2007}{224101}.

\bibitem{x9}
\Name{Wang W.-X., Yang R., Lai Y.-C., Kovanis V. \and Grebogi C.} \REVIEW{Phys.
  Rev. Lett.}{106}{2011}{154101}.

\bibitem{x10}
\Name{Levnaji{\'c} Z. \and Pikovsky A.} \REVIEW{Phys. Rev.
  Lett.}{107}{2011}{034101}.

\bibitem{x11}
\Name{Yu D., Righero M. \and Kocarev L.} \REVIEW{Phys. Rev.
  Lett.}{97}{2006}{188701}.

\bibitem{x12}
\Name{Geier F., Timmer J. \and Fleck C.} \REVIEW{BMC Systems
  Biology}{1}{2007}{11}.

\bibitem{x13}
\Name{Hempel S., Koseska A., Kurths J. \and Nikoloski Z.} \REVIEW{Phys. Rev.
  Lett.}{107}{2011}{054101}.

\bibitem{x6}
\Name{Shandilya S.~G. \and Timme M.} \REVIEW{New J. Phys.}{13}{2011}{013004}.

\bibitem{x7}
\Name{Strogatz S.~H.} \REVIEW{Nature}{410}{2001}{268}.

\bibitem{x18}
\Name{Yang P., Li X., Wu M., Kwoh C.-K. \and Ng S.-K.} \REVIEW{PLoS
  ONE}{6}{2011}{e21502}.

\bibitem{x14}
\Name{Murphy K.~P.} \Book{Dynamic bayesian networks} Ph.D. thesis UC Berkeley
  (2002).

\bibitem{x19}
\Name{Maucher M., Kracher B., K{\"u}hl M. \and Kestler H.~A.}
  \REVIEW{Bioinformatics}{27}{2011}{1529}.

\bibitem{x16}
\Name{Wang W.-X., Yang R., Lai Y.-C., Kovanis V. \and Harrison M. A.~F.}
  \REVIEW{Europhys. Lett.}{94}{2011}{48006}.

\bibitem{x15}
\Name{Roudi Y. \and Hertz J.} \REVIEW{Phys. Rev. Lett.}{106}{2011}{048702}.

\bibitem{x17}
\Name{Yu D. \and Parlitz U.} \REVIEW{PLoS ONE}{6}{2011}{e24333}.

\bibitem{altarelli2014bayesian}
\Name{Altarelli F., Braunstein A., Dall¡¯Asta L., Lage-Castellanos A. \and
  Zecchina R.} \REVIEW{Phys. Rev. Lett.}{112}{2014}{118701}.

\bibitem{liao2015reconstructing}
\Name{Liao H. \and Zeng A.} \REVIEW{Sci. Rep.}{5}{2015}{11404}.

\bibitem{x20}
\Name{Wells K., O¡¯Hara R.~B., Pfeiffer M., Lakim M.~B., Petney T.~N. \and
  Durden L.~A.} \REVIEW{Oecologia}{172}{2013}{307}.

\bibitem{x21}
\Name{Barreiro M., Marti A.~C. \and Masoller C.}
  \REVIEW{Chaos}{21}{2011}{013101}.

\bibitem{x22}
\Name{Kim M. \and Leskovec J.} \Book{The Network Completion Problem: Inferring
  Missing Nodes and Edges in Networks} Ch.~5 pp. 47--58.

\bibitem{Guimera29122009}
\Name{Guimer{\`a} R. \and Sales-Pardo M.} \REVIEW{Proc. Natl. Acad. Sci.
  USA}{106}{2009}{22073}.

\bibitem{Lu20111150}
\Name{L{\"u} L. \and Zhou T.} \REVIEW{Physica A}{390}{2011}{1150}.

\bibitem{Lv20121}
\Name{L{\"u} L., Medo M., Yeung C.~H., Zhang Y.-C., Zhang Z.-K. \and Zhou T.}
  \REVIEW{Phys. Rep.}{519}{2012}{1}.

\bibitem{PhysRevLett.92.118701}
\Name{Noh J.~D. \and Rieger H.} \REVIEW{Phys. Rev. Lett.}{92}{2004}{118701}.

\bibitem{Wittmann20093826}
\Name{Wittmann D.~M., Schmidl D., Bl{\"o}chl F. \and Theis F.~J.}
  \REVIEW{Theoretical Computer Science}{410}{2009}{3826}.

\bibitem{note}
We think that the network we will reconstruct is connected, so there are at
  least $N-1$ links in the network.

\bibitem{LusseauS477}
\Name{Lusseau D. \and Newman M. E.~J.} \REVIEW{Proceedings of the Royal Society
  of London B: Biological Sciences}{271}{2004}{S477}.

\bibitem{wang2003epidemic}
\Name{Wang Y., Chakrabarti D., Wang C. \and Faloutsos C.} \Book{Epidemic
  spreading in real networks: an eigenvalue viewpoint} in proc. of
  \Book{Reliable Distributed Systems, 2003. Proceedings. 22nd International
  Symposium on} 2003 pp. 25--34.

\end{thebibliography}

\end{document}